\documentclass[conference]{IEEEtran}
\IEEEoverridecommandlockouts
\usepackage{cite}
\usepackage{amsmath,amssymb,amsfonts}
\usepackage{amsthm}
\newtheorem{example}{Example}
\usepackage{algorithmic}
\usepackage{graphicx}
\usepackage{textcomp}
\usepackage{xcolor}
\usepackage{enumitem}
\usepackage{cleveref}
\usepackage{enumitem}
\usepackage{soul}
\usepackage{caption, subcaption}
\usepackage{xcolor,listings}
\usepackage[acronym]{glossaries}
\usepackage{enumitem}
\usepackage{cleveref}
\usepackage{listings}
\usepackage{xcolor}
\usepackage{booktabs}
\usepackage{url}

\lstdefinelanguage{json}{
    basicstyle={\footnotesize\ttfamily},
    numbers=left,
    numberstyle=\tiny\color{gray},
    stepnumber=1,
    numbersep=5pt,
    showstringspaces=false,
    breaklines=true,
    frame=lines,
    keywordstyle=\color{blue},
    stringstyle=\color{green!40!black},
    moredelim=[s][\color{black}]{\{}{\}},
    moredelim=[s][\color{black}]{[}{]},
    literate=
      *{:}{{\color{black}:}}{1}%
       {,}{{\color{black},}}{1}%
       {"}{{\textquotedbl}}{1}%
       {0}{{\color{black}0}}{1}%
       {1}{{\color{black}1}}{1}%
       {2}{{\color{black}2}}{1}%
       {3}{{\color{black}3}}{1}%
       {4}{{\color{black}4}}{1}%
       {5}{{\color{black}5}}{1}%
       {6}{{\color{black}6}}{1}%
       {7}{{\color{black}7}}{1}%
       {8}{{\color{black}8}}{1}%
       {9}{{\color{black}9}}{1}%
}

\def\BibTeX{{\rm B\kern-.05em{\sc i\kern-.025em b}\kern-.08em
    T\kern-.1667em\lower.7ex\hbox{E}\kern-.125emX}}
\begin{document}

\title{Query Performance Explanation through Large Language Model for HTAP Systems
% \thanks{Identify applicable funding agency here. If None, delete this.}
}

\author{ \IEEEauthorblockN{Haibo Xiu\IEEEauthorrefmark{1}
% \thanks{* Work completed during internship at ByteDance US (TikTok), San Jose, CA, USA.}
, Li Zhang\IEEEauthorrefmark{2}, Tieying Zhang\IEEEauthorrefmark{2}, Jun Yang\IEEEauthorrefmark{1}, Jianjun Chen\IEEEauthorrefmark{2}}
    \IEEEauthorblockA{\IEEEauthorrefmark{1}\textit{Duke University}, Durham, NC, USA \\
    }
    \IEEEauthorblockA{\IEEEauthorrefmark{2}\textit{US Infrastructure System Lab}, ByteDance, Inc \\
    haibo.xiu@duke.edu, \{li.zhang, tieying.zhang, jianjun.chen\}@bytedance.com, junyang@cs.duke.edu}

% \and
% \and
% \and
% \and
% \and
}

\maketitle

\begin{abstract}
In hybrid transactional and analytical processing (HTAP) systems, users often struggle to understand why query plans from one engine (OLAP or OLTP) perform significantly slower than those from another. Although optimizers provide plan details via the \texttt{EXPLAIN} function, these explanations are frequently too technical for non-experts and offer limited insights into performance differences across engines. To address this, we propose a novel framework that leverages large language models (LLMs) to explain query performance in HTAP systems. 
Built on Retrieval-Augmented Generation (RAG), our framework constructs a knowledge base that stores historical query executions and expert-curated explanations. 
To enable efficient retrieval of relevant knowledge, query plans are embedded using a lightweight tree-CNN classifier. 
This augmentation allows the LLM to generate clear, context-aware explanations of performance differences between engines. 
Our approach demonstrates the potential of LLMs in hybrid engine systems, paving the way for further advancements in database optimization and user support.
\end{abstract}


\section{Introduction}
\textcolor{red}{
% TODO: \\
% 1. add a note, we are explaining the reason of why one engine is faster than the other, instead of why our system chooses this engine to execute. \\
% 2. mention that, we are not going to many details about the tree-CNN classifier here, since that is included in the sigmod 2025 paper.\\
}
``Why does my query run so slowly?'' In modern database management systems (DBMS), users frequently struggle to understand why certain queries experience very long execution times. While contemporary optimizers provide an \texttt{EXPLAIN} function that details the execution plan, these explanations are still too complex for non-experts to interpret fully. This challenge is particularly evident in hybrid transactional and analytical processing (HTAP) systems, such as ByteHTAP\cite{chen2022bytehtap} developed at ByteDance, which features a unified interface with two underlying execution engines: OLTP (online transactional processing, referred to as TP) and OLAP (analytical processing, referred to as AP). When a query arrives, users often need guidance on selecting the optimal engine and understanding why one engine might perform better than the other. Traditionally, database experts manually analyze queries to provide tailored explanations, but as query volumes grow, this approach becomes unsustainable.

To fill the gap between the incomprehensibility of optimizer-generated explanations and the high cost of expert-provided explanations, we aim to develop a user-friendly and intelligent explanation system satisfying three key criteria:
\begin{itemize}[leftmargin=*]
    \item The explanations should be clear and easy for non-experts to understand.
    \item The explanations must consider the database and engine context to ensure reasonably accurate output, even if they may not always achieve the same level of precision as human experts.
    \item The solution must be cost-effective in terms of training and maintenance.
\end{itemize}

To meet these requirements, we propose a novel framework that leverages large language models (LLMs) to automatically explain query performance across different engines. LLMs have gained popularity thanks to their ability to generate understandable natural language outputs. However, balancing accuracy and efficiency presents a trade-off: while using pre-trained models (such as Doubao\cite{Doubao}, ChatGPT\cite{achiam2023gpt}, Llama\cite{touvron2023llama}, and Claude\cite{bai2022constitutional}) is efficient, they lack the specific context needed for accurate query performance explanations. 
Although fine-tuning LLMs could improve relevance and accuracy, this is resource-intensive. 
To strike a balance between cost and accuracy, we employ pre-trained public models augmented by a Retrieval-Augmented Generation (RAG) approach\cite{rag2020}, addressing the limitations of general-purpose LLMs while maintaining efficiency.

The RAG framework relies on two key components: a retriever and a knowledge base\cite{rag2023}. The retriever finds relevant references in a pre-built knowledge base and provide them as contextual input to the LLM. In the HTAP system ByteHTAP\cite{chen2024bytehtap} that this paper focuses on, this retriever is powered by a lightweight machine learning model---specifically, a tree-CNN classifier based on recent research on learned query optimizers\cite{marcus2021bao, zhu2023lero, xu2023coool}, trained to route queries to the engine best suited for efficient execution. 
This model also functions as a query plan encoder, generating embeddings that represent the plan for each query.
The knowledge base serves as an external data source, supplementing the LLM's original training data. In our setup, the knowledge base stores historical queries, their plan embeddings generated by the tree-CNN classifier, and expert-curated performance difference explanations, i.e., reasons why a query runs faster on AP or TP.\footnote{These explanations address why one engine's plan runs faster than the other post-execution, rather than merely interpreting the classifier's routing decision---the latter underscores a broader challenge of interpretability in machine learning.}
At runtime, the retriever searches the knowledge base for similar plan embeddings relevant to the current query. These retrieved embeddings enrich the LLM with relevant context-specific information, enabling it to generate more accurate and targeted natural language explanations. 
By combining the precision of the machine learning model's embeddings with the LLM's generative capabilities, experiments show that our approach provides insightful, contextually grounded explanations of engine performance, without requiring expert intervention or costly fine-tuning of the LLM. 
Our framework is not limited to ByteHTAP; it is adaptable to any HTAP system capable of retrieving historical explanations as augmented inputs.

Explaining why one engine outperforms another may seem more straightforward than answering the broader question of ``Why does my query run so slowly''. However, the two are intricately connected, as engine choice and query execution performance are shaped by the same underlying factors, such as plan efficiency and system architecture. While fully automating explanations for query performance remains challenging, our approach offers a significant step forward. By integrating LLMs within HTAP systems, our framework helps users understand engine performance differences, demonstrates the potential of LLMs to deliver intuitive, accessible explanations, and paves way for future research toward more comprehensive automation in database performance analysis.

The structure of this paper is as follows: After briefly surveying related work in \Cref{sec:related},
we describe our framework in \Cref{sec:our-method}, including the integration of LLMs with RAG to improve explanation quality in HTAP systems. \Cref{sec:knowledge-base} and \Cref{sec:prompt} outlines the knowledge base construction and prompt engineering, while \Cref{sec:exp} presents and experimental analyzes the results. Finally, \Cref{sec:future} concludes with key insights and future research directions.

\section{Related work}\label{sec:related}
Large Language Models (LLMs) have shown considerable promise in enhancing database management by providing intuitive, natural language explanations and recommendations \cite{zhou2024llm}. LLMs excel at translating complex, system-level insights into user-friendly explanations, sparking interest in their application across various database tasks, including query performance analysis and optimization. LLMs have also been effectively used for database diagnostics; for instance, D-Bot \cite{zhou2023d} employs LLMs to detect and resolve anomalies within databases, while Panda \cite{singh2024panda} leverages Retrieval-Augmented Generation (RAG)\cite{lewis2020retrieval} to ground LLM outputs in context, providing performance diagnostics based on execution metrics rather than query plan analysis.
RAG improves the accuracy and relevance of language models by integrating a retrieval mechanism with text generation\cite{rag2023}.
Further research, such as DBG-PT \cite{DBG-PT}, demonstrates the utility of LLMs for diagnosing performance regressions through comparisons between structured query plans. 
While DBG-PT successfully analyzes plans from the same optimizer, our work extends this approach to scenarios involving plans generated by different engines. Instead of comparing the plan details from the \texttt{EXPLAIN} clause directly, our methods enhances explanation quality by integrating RAG by a small model for improved context and relevance. 
% Similar idea of combining 

% Topics: LLM, LLM4DB, apply LLM to explain, learned optimizer, DBG-PT, Pandas

\section{Our method: retrieval-augmented explanation generation by Large Language Model}\label{sec:our-method}
% \subsection{Motivation for Using RAG}
 % In our approach, the retriever searches a knowledge base containing historical queries and expert explanations to identify similar plan encodings. 
\subsection{Constructing an Effective Retriever}
Since LLMs may lack query-specific or up-to-date context, we leverage insights from past queries through RAG to enrich responses with precise and relevant historical information. These retrieved references, combined with carefully designed prompts, are provided to the LLM to generate more accurate and contextually grounded answers.
\paragraph{Knowledge base for RAG}
In our system, we store historical query plans and their corresponding plan performance explanations in a key-value knowledge base. 
The key consists of a pair of query plans, while the value contains the plan details and associated explanations. 
Instead of straightforwardly storing the plan pairs as raw text (e.g., the output of \texttt{EXPLAIN} from the optimizer), the plan pairs in our system are stored as vectors (the encoding process is described later).
This idea is motivated by our focus not on semantic similarity between plans but on ``similar performance distinctions''—specifically, similar performance differences between TP and AP plans observed in past queries.
% For a new query, we search the knowledge base and retrieve the top K most similar plan vectors, measuring similarity using metrics like Euclidean distance between encoding vectors.
For a new query, we aim to retrieve historical queries with similar plan performance distinctions and use this knowledge to enhance the LLM's generation.

\paragraph{Plan embeddings by a lightweight model}
Our HTAP system\cite{chen2024bytehtap} features a smart router, which is an enhanced tree-CNN classifier that predicts, for a given query, which engine will yield a plan with better performance. Experiments demonstrated that the router achieves high accuracy in identifying the more efficient plan between TP and AP engines. Therefore, it naturally serves as a good model for generating plan embeddings.
Key advantages of using the smart router for plan embeddings include:
\begin{itemize}[leftmargin=*]
    \item \textit{Lightweight model:} The smart router is highly efficient, with a physical model size of less than 1MB and an average inference time of only 1ms, making it an ideal choice for embedding generation. Additionally, it can be quickly retrained to adjust to changes in query workloads or underlying data.
    \item \textit{Task-specific design:} 
    Directly taking plan trees as input, embeddings generated by the smart router can capture detailed plan performance comparisons since the original task is to determine the faster engine. Trained on a large dataset of query plan pairs, it identifies performance-relevant features within plans. These intermediate plan encodings serve as ``identifiers'', enabling the retriever to match new queries with similar historical performance distinctions. 
\end{itemize}

\subsection{Framework Overview}\label{sec:framework}

\begin{figure*}[t]
    \centering
    \includegraphics[scale=0.4]{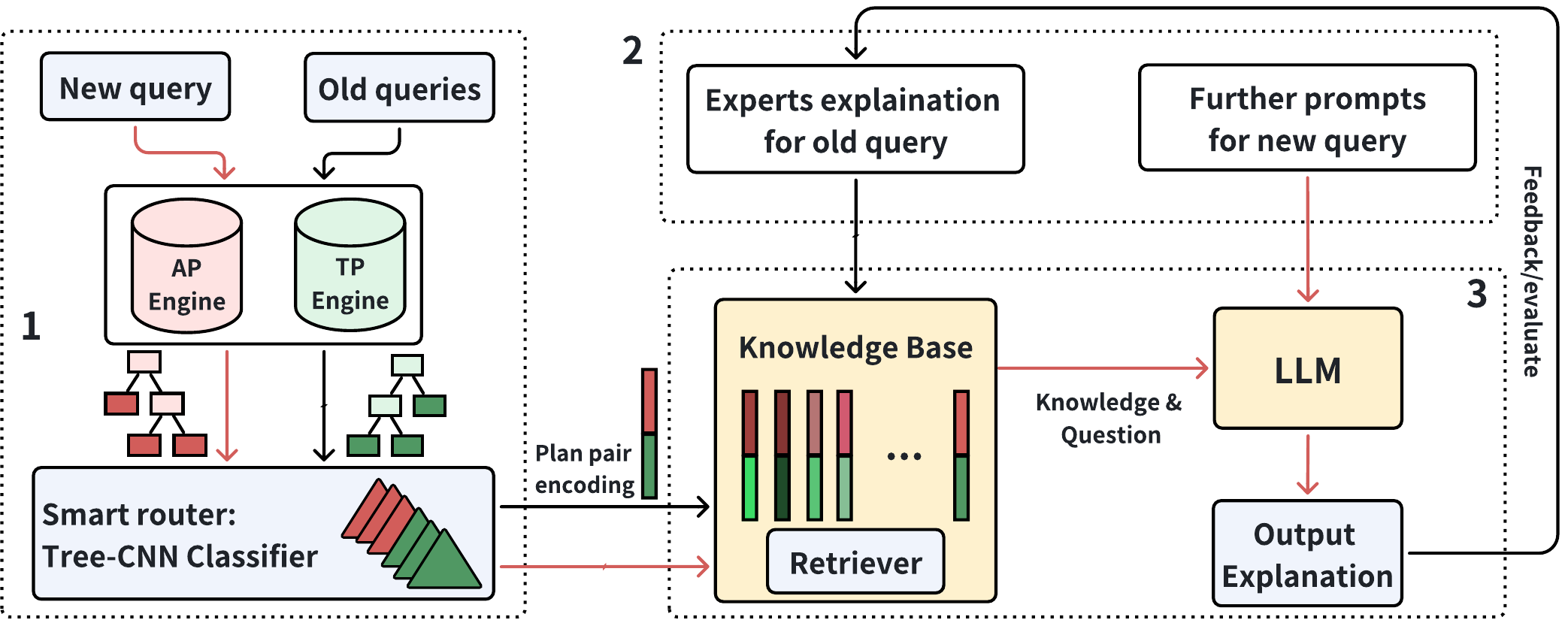}
    \caption{
Framework of our method with three main components: \textbf{1}: HTAP system, \textbf{2}: Human (users/experts) side, and \textbf{3}: RAG and LLM side. The red line shows the workflow for new queries, while the black line represents historical queries.
}
    \label{fig:framework}
    % \vspace{-2mm}
\end{figure*}

From a system integration perspective, our explainer operates above the TP/AP optimizer in the ByteHTAP system. The system steers a pre-trained LLM to generate explanations based on plan embeddings from the smart router and retrieved contextual information. The framework consists of three main components
% : HTAP system with smart router side, human side, and the RAG-based LLM explainer side,
as illustrated in \Cref{fig:framework}, which we describe below in turn.

\paragraph{ByteHTAP system with smart router} This component is marked as \textbf{$\mathbf{1}$} in \Cref{fig:framework}. 
% queries are routed to either AP or TP engine based on the smart router's prediction. 
To explain the performance distinction between engines for a new query (marked by red arrows), the tree-structured execution plans from both AP and TP engines are processed by the smart router, which encodes them into a vector representation. The plan pair embedding, created by concatenating vectors from both AP and TP plans, is then utilized by subsequent components to generate explanations.
Historical queries (marked by black arrows) are selected from the training set of smart router and forwarded to the human side for expert-curated explanations, which are then stored in the knowledge base along with their corresponding plan pair embeddings. 
% Further details about knowledge base construction are in \Cref{sec:knowledge-base}. 

\paragraph{Human interaction and expert evaluation}
The human side (marked as \textbf{$\mathbf{2}$}) has two roles: database users and database experts. Users submit queries and seek guidance on performance-related questions. Experts can provide detailed explanations of why one plan performs better or worse than the other based on practical insights. They can also assess the quality of future LLM-generated explanations. Users may also offer additional contextual information, such as details on newly created indexes, which helps refine the LLM's responses and improve explanation accuracy.

\paragraph{RAG and LLM integration} 
On this side (marked as \textbf{$\mathbf{3}$} in \Cref{fig:framework}), we integrate a knowledge base for retrieval and a pre-trained public LLM for explanation generation.
The knowledge base is a vector database populated with historical queries, where their AP/TP execution plans are encoded by the smart router.
The resulting plan pair embeddings are stored as keys and the values are the expert's explanations.
Further details on the construction of the knowledge base are provided in \Cref{sec:knowledge-base}. 
For each incoming user query, the retriever uses the plan pair embedding obtained from the smart router to search in the knowledge base for the top $K$ most similar plan pairs.
These retrieved knowledge, along with expert explanations, background context, and any user-provided prompts, serve as augmented input for the LLM. The LLM then generates an explanation based on this enriched input, which is returned to the user.\footnote{If the LLM determines the augmented knowledge lacks sufficient information, it will return a None response.} As mentioned, these generated outputs can also be reviewed and evaluated by experts. If an explanation is determined inaccurate, experts will correct it and add the revised version to the knowledge base for future retrieval.

\section{RAG Knowledge Base Construction}\label{sec:knowledge-base}
For RAG, we construct a knowledge base by storing historical queries and their performance explanations. This knowledge base provides the necessary context for the LLM to generate accurate and relevant explanations. For each query, we store the following data:
\textbf{$<$}\textit{plan pair encoding,} \textit{plan details}, \textit{execution result}, \textit{expert explanation}$>$.
The plan pair encoding is a vectorized representation of the pair of AP and TP plans, encoded by the smart router, which enables efficient retrieval of similar queries. Plan details includes the actual execution plans for both engines. The execution result indicates which engine executes this query faster. Finally, the expert explanation is a curated explanation from database experts detailing why one engine outperforms the other in specific cases. 

To ensure that the knowledge base represents common queries, we select query patterns frequently requested by users seeking explanations for why one engine is slower than the other. Furthermore, we synthetize similar queries using the TPC-H 
% (100GB) 
schema. Note that these generated queries are also in the training set of the smart router, ensuring the encodings are attended to the performance distinctions. These query patterns primarily include:
\begin{enumerate} 
\item \textit{Join queries}: Join operations in which AP and TP engines apply different join strategies, offering insights into engine-specific optimizations. These join queries vary in factors such as the number of joined tables, table size, predicate selectivity, and index usage. 
\item \textit{Top-$N$ queries}: Queries that retrieve the top $N$ records based on specific criteria, often using clauses like \texttt{ORDER BY}, \texttt{LIMIT}, and sometimes \texttt{OFFSET}. These queries are common in user workloads and often perform differently depending on engine-specific optimizations.
\end{enumerate}
After executing these synthetic queries on both TP and AP engines, we send the queries, plans, and execution results to database experts, requesting the corresponding explanations. % To keep vector search times manageable
For the experimental setup in \Cref{sec:exp}, we selectively include only 20 representative queries in the knowledge base.
On one hand, this limited number of queries helps reduce the cost of expert annotations. 
On the other hand, we hypothesize that this small set is sufficient to capture the performance distinctions that are broadly applicable to the users' query workload.
For each query, the plan pair encoding is a 16-dim vector.
The retriever searches the top 2 similar vectors for the new query.
An additional 200 synthetic queries are used as the test set. We also provide the interface for the knowledge base to accept new queries with experts explanations. 
A full analysis of how to manage the knowledge base, including methods for automatically selecting representative queries and expiring stale queries, remains a future work.

\section{Prompt Engineering}\label{sec:prompt}
Following the RAG process, we send the retrieved information along with the new query as to the LLM. To guide the LLM in generating accurate explanations, we provide carefully structured prompts, which are organized into three parts:
\begin{itemize}[leftmargin=*] 
\item \textit{Background information}, which includes the overall objective, an overview of the HTAP system, key differences between its engines, and specifics about the schema and dataset queried.
\item \textit{Task description}, which defines the RAG task, detailing expected inputs and outputs, along with additional guidance to ensure clarity.
\item \textit{Additional user-provided context}, such as recent modifications to indexes, to ensure that the LLM has the latest context.
\end{itemize}

During prompt design, we observed that the pre-trained LLM often defaults to directly comparing the plan costs generated by the query optimizer to explain which plan is faster.
However, because the optimizers are implemented differently across engines, plan costs are generally not comparable across engines (which is the case for ByteHTAP). To prevent this incorrect reasoning, we emphasized in the prompts that the costs in the plan pair should not be used for comparison. \Cref{tab:prompt} presents the prompts used in our experiments, and \Cref{sec:exp} analyzes the generation results.

\begin{table}[t]
    \centering
    \footnotesize
    \begin{tabular}{p{8cm}}
    \toprule
    \textbf{Background information: }\footnotesize\textit{We are using RAG to assist database users in understanding query performance across differences engines in our HTAP system—specifically, why one engine performs faster while the other is slower.
Please ensure you are familiar with the TPC-H schema, and our dataset follows the default schema and contains 100GB of data.
Our HTAP system has two database engines, ``TP'' and ``AP''. The TP engine uses row-oriented storage, while the AP engine utilizes column-oriented storage.
Note that the optimizers for TP and AP engines are distinct, leading to different execution plans. Therefore, you are not allowed to compare the cost estimates of the execution plans from TP and AP engines. }\\
    \midrule
\textbf{Task description: }\textit{Here is your task: I will input you the execution plans for the query from both the TP and AP engines, please evaluate the likely performance of each engine without directly comparing the cost estimates. Focus on factors such as the join methods used, the storage formats (row-oriented vs. column-oriented), index utilization, and any potential implications of the execution plan characteristics on query performance. Your task is to explain which engine might perform better for this specific query and why, based on these factors.} \textit{To assist you, we have a retriever that can find relevant historical plans from the knowledge base with precise performance explanation from our experts.
The \textbf{KNOWLEDGE} and \textbf{QUESTIONS} you received will be in the following format:}

\begin{itemize}[leftmargin=*]
\item \textit{\textbf{KNOWLEDGE}:  historical query + historical plan pair (AP/TP's plan) + historical execution result (indicating whether TP or AP is faster) + historical expert explanation (why TP or AP is faster).}
    \item \textit{\textbf{QUESTION}: new query + new plan pair + new execution result.}
\end{itemize}
\textit{You could use {KNOWLEDGE} to explain the following new pair of plans in {QUESTION}.  If the {KNOWLEDGE} does not contain the facts to answer the {QUESTION} return {None}. Note, to make sure your answer is accurate, I may input you several retrieved old queries with their plans, results and explanations. Please understand all the information I provide to generate your explanation. Now, I am ready to send you the KNOWLEDGE and QUESTION}
\\
\midrule
\textbf{Additional user context:}
\footnotesize
\textit{Beyond the default indexes on primary and foreign keys, an additional index has been created on the \texttt{c\_phone} column in the \texttt{customer} table. }\\
\bottomrule
    \caption{\small Prompt engineering}
    \label{tab:prompt}
    % \vspace{-3mm}
    \end{tabular}
\end{table}

\section{Experimental Analysis and Participant Study}\label{sec:exp}

\subsection{Demonstrative Case}
First, we show one example to demonstrate how our system works. This query is synthetic after we observe similar query patterns from the real users queries. 
All subsequent queries were executed in the same environment described in \cite{chen2024bytehtap}, consisting of a six-machine cluster with four data servers. Each data server is configured with 8 vCPUs, 32 GB DRAM, and 1 NUMA node. The nodes utilize unified storage, run Debian 4.14.81, and are connected via a 25 Gbps Ethernet network.

\begin{example}\label{eg:q3}
    Consider a query joining 3 tables. TP's plan takes $5.80$s to run, while AP's plan completes in $310$ms. We show the details of TP and AP plans in \Cref{tab:plan} and the corresponding expert's explanation and the LLM-generated explanation in \Cref{tab:explain}.
\vspace*{1.5ex}
\begin{lstlisting}[language=SQL, basicstyle=\footnotesize\ttfamily]
SELECT COUNT(*) FROM customer, nation, orders         
WHERE SUBSTRING(c_phone, 1, 2) IN ('20', '40', '22', '30', '39', '42', '21') 
AND c_mktsegment = 'machinery'   
AND n_name = 'egypt' AND o_orderstatus = 'p'           
AND o_custkey = c_custkey 
AND n_nationkey = c_nationkey;   
\end{lstlisting}
\end{example}

In this example, the explanation generated by our approach using the LLM demonstrates high accuracy. It highlights the key factor that hash joins are more efficient than nested-loop joins, as no index is available, which aligns closely with the expert explanation. The LLM-generated explanation also providing additional insights, including details about AP's aggregation efficiency, an aspect the experts did not explicitly mention. Overall, LLM output is informative, clear, and easy to understand by non-experts. Since the ease of understanding is a subjective measure, we conducted a user study to gather feedback on how well users comprehended the generated explanations. The design and results of the user study are detailed in \Cref{sec:study}.

\begin{table}[]
    \centering
    \footnotesize
    \begin{tabular}{p{8cm}}
    \toprule
    \textbf{Details of TP's Plan for \Cref{eg:q3}} \\
    \midrule
        \footnotesize{\{
    \textquotesingle \textit{Node Type}\textquotesingle : \textquotesingle Group aggregate\textquotesingle,
    \textquotesingle \textit{Total Cost}\textquotesingle : 5213.0,
    \textquotesingle \textit{Plan Rows}\textquotesingle : 1,
    \textquotesingle \textit{Plans}\textquotesingle : [
        \{
            \textquotesingle \textit{Node Type}\textquotesingle : \textquotesingle Nested loop inner join\textquotesingle,
            \textquotesingle \textit{Total Cost}\textquotesingle : 5175.0,
            \textquotesingle \textit{Plan Rows}\textquotesingle : 379,
            \textquotesingle \textit{Plans}\textquotesingle : [
                \{
                    \textquotesingle \textit{Node Type}\textquotesingle : \textquotesingle Nested loop inner join\textquotesingle,
                    \textquotesingle \textit{Total Cost}\textquotesingle : 1002.0,
                    \textquotesingle \textit{Plan Rows}\textquotesingle : 285,
                    \textquotesingle \textit{Plans}\textquotesingle : [
                        \{
                            \textquotesingle \textit{Node Type}\textquotesingle : \textquotesingle Filter\textquotesingle,
                            \textquotesingle \textit{Total Cost}\textquotesingle : 2.75,
                            \textquotesingle \textit{Plan Rows}\textquotesingle : 2,
                            \textquotesingle \textit{Plans}\textquotesingle : [
                                \{
                                    \textquotesingle \textit{Node Type}\textquotesingle : \textquotesingle Table Scan\textquotesingle,
                                    \textquotesingle \textit{Relation Name}\textquotesingle : \textquotesingle nation\textquotesingle,
                                    \textquotesingle \textit{Total Cost}\textquotesingle : 2.75,
                                    \textquotesingle \textit{Plan Rows}\textquotesingle : 25
                                \}
                            ]
                        \},
                        \{
                            \textquotesingle \textit{Node Type}\textquotesingle : \textquotesingle Filter\textquotesingle,
                            \textquotesingle \textit{Total Cost}\textquotesingle : 290.0,
                            \textquotesingle \textit{Plan Rows}\textquotesingle : 114,
                            \textquotesingle \textit{Plans}\textquotesingle : [
                                \{
                                    \textquotesingle \textit{Node Type}\textquotesingle : \textquotesingle Table Scan\textquotesingle,
                                    \textquotesingle \textit{Relation Name}\textquotesingle : \textquotesingle customer\textquotesingle,
                                    \textquotesingle \textit{Total Cost}\textquotesingle : 290.0,
                                    \textquotesingle \textit{Plan Rows}\textquotesingle : 1142
                                \}
                            ]
                        \}
                    ]
                \},
                \{
                    \textquotesingle \textit{Node Type}\textquotesingle : \textquotesingle Filter\textquotesingle,
                    \textquotesingle \textit{Total Cost}\textquotesingle : 13.3,
                    \textquotesingle \textit{Plan Rows}\textquotesingle : 1,
                    \textquotesingle \textit{Plans}\textquotesingle : [
                        \{
                            \textquotesingle \textit{Node Type}\textquotesingle : \textquotesingle Table Scan\textquotesingle,
                            \textquotesingle \textit{Relation Name}\textquotesingle : \textquotesingle orders\textquotesingle,
                            \textquotesingle \textit{Total Cost}\textquotesingle : 13.3,
                            \textquotesingle \textit{Plan Rows}\textquotesingle : 13
                        \}
                    ]
                \}
            ]
        \}
    ]
\}
}
\\
\midrule
\textbf{Details of AP's plan for \Cref{eg:q3}}\\
\midrule
\footnotesize{\{
    \textquotesingle \textit{Node Type}\textquotesingle : \textquotesingle Aggregate\textquotesingle,
    \textquotesingle \textit{Total Cost}\textquotesingle : 16500000.0,
    \textquotesingle \textit{Plan Rows}\textquotesingle : 1,
    \textquotesingle \textit{Plans}\textquotesingle : [
        \{
            \textquotesingle \textit{Node Type}\textquotesingle : \textquotesingle Inner hash join\textquotesingle,
            \textquotesingle \textit{Total Cost}\textquotesingle : 16500000.0,
            \textquotesingle \textit{Plan Rows}\textquotesingle : 134933,
            \textquotesingle \textit{Plans}\textquotesingle : [
                \{
                    \textquotesingle \textit{Node Type}\textquotesingle : \textquotesingle Filter\textquotesingle,
                    \textquotesingle \textit{Total Cost}\textquotesingle : 13500000.0,
                    \textquotesingle \textit{Plan Rows}\textquotesingle : 13500000,
                    \textquotesingle \textit{Plans}\textquotesingle : [
                        \{
                            \textquotesingle \textit{Node Type}\textquotesingle : \textquotesingle Table Scan\textquotesingle,
                            \textquotesingle \textit{Relation Name}\textquotesingle : \textquotesingle orders\textquotesingle,
                            \textquotesingle \textit{Total Cost}\textquotesingle : 0.5,
                            \textquotesingle \textit{Plan Rows}\textquotesingle : 135000000
                        \}
                    ]
                \},
                \{
                    \textquotesingle \textit{Node Type}\textquotesingle : \textquotesingle Hash\textquotesingle,
                    \textquotesingle \textit{Plans}\textquotesingle : [
                        \{
                            \textquotesingle \textit{Node Type}\textquotesingle : \textquotesingle Inner hash join\textquotesingle,
                            \textquotesingle \textit{Total Cost}\textquotesingle : 1640000.0,
                            \textquotesingle \textit{Plan Rows}\textquotesingle : 135985,
                            \textquotesingle \textit{Plans}\textquotesingle : [
                                \{
                                    \textquotesingle \textit{Node Type}\textquotesingle : \textquotesingle Filter\textquotesingle,
                                    \textquotesingle \textit{Total Cost}\textquotesingle : 1500000.0,
                                    \textquotesingle \textit{Plan Rows}\textquotesingle : 1360000,
                                    \textquotesingle \textit{Plans}\textquotesingle : [
                                        \{
                                            \textquotesingle \textit{Node Type}\textquotesingle : \textquotesingle Table Scan\textquotesingle,
                                            \textquotesingle \textit{Relation Name}\textquotesingle : \textquotesingle customer\textquotesingle,
                                            \textquotesingle \textit{Total Cost}\textquotesingle : 0.5,
                                            \textquotesingle \textit{Plan Rows}\textquotesingle : 13600000
                                        \}
                                    ]
                                \},
                                \{
                                    \textquotesingle \textit{Node Type}\textquotesingle : \textquotesingle Hash\textquotesingle,
                                    \textquotesingle \textit{Plans}\textquotesingle : [
                                        \{
                                            \textquotesingle \textit{Node Type}\textquotesingle : \textquotesingle Filter\textquotesingle,
                                            \textquotesingle \textit{Total Cost}\textquotesingle : 3.0,
                                            \textquotesingle \textit{Plan Rows}\textquotesingle : 2,
                                            \textquotesingle \textit{Plans}\textquotesingle : [
                                                \{
                                                    \textquotesingle \textit{Node Type}\textquotesingle : \textquotesingle Table Scan\textquotesingle,
                                                    \textquotesingle \textit{Relation Name}\textquotesingle : \textquotesingle nation\textquotesingle,
                                                    \textquotesingle \textit{Total Cost}\textquotesingle : 0.5,
                                                    \textquotesingle \textit{Plan Rows}\textquotesingle : 25
                                                \}
                                            ]
                                        \}
                                    ]
                                \}
                            ]
                        \}
                    ]
                \}
            ]
        \}
    ]
\}
}\\
\bottomrule
    \caption{\small TP and AP's plan for \Cref{eg:q3}.}
    % \vspace{-2mm}
    
    \label{tab:plan}
    \end{tabular}
\end{table}

\begin{table}[t]
    \centering
    \footnotesize
    \begin{tabular}{p{8cm}}
    \toprule
    \textbf{Explanation by experts for \Cref{eg:q3}}\\
    \midrule
\footnotesize{\textit{AP is faster than TP because TP has to use nested loop join with no index available. AP uses hash join, which is more efficient.}}\\
    \midrule
\textbf{Explanation by our approach for \Cref{eg:q3}}\\
\midrule
\footnotesize{\textit{AP is faster due to its use of hash joins and hash aggregates, which are highly efficient for handling large datasets, especially in a columnar storage format. These operations allow AP to scan only relevant columns and apply filters before joining, making it particularly effective for large, complex queries involving multiple tables. In contrast, TP's use of nested loop joins and group aggregates, combined with table scans that don't benefit from index optimizations, leads to slower performance when processing large datasets. AP's ability to efficiently process joins and aggregates across multiple tables gives it a significant advantage in this scenario.}}\\
\midrule
\textbf{Explanation by DBG-PT for \Cref{eg:q3}}\\
\midrule
\footnotesize{\textit{The AP engine is faster in this case because it utilizes column-oriented storage, which efficiently scans large tables like orders by only reading the required columns. Additionally, the AP engine uses hash joins, which are well-suited for joining large datasets. Both engines likely benefit from the index on \texttt{c\_phone}; however, the AP engine's column-oriented storage allows it to quickly access and filter the \texttt{c\_phone} column without the overhead of retrieving entire rows. In contrast, the TP engine performs nested loop joins that are less efficient for large tables and requires reading entire rows, making table scans more costly. These factors combined result in the AP engine delivering better performance for this query.}}\\
\bottomrule
    \caption{\small Plan performance explanations for \Cref{eg:q3} by different approaches.}
    \label{tab:explain}
    % \vspace{-2mm}
    \end{tabular}
\end{table}

\subsection{Experimental analysis}
We first focus on experimentally evaluating the accuracy of the LLM-generated explanations.
Given the lack of a standard evaluation method or benchmark for this specific setting, we relied on three database experts on HTAP systems to manually assess the generated explanations for correctness and completeness.
The experts' evaluations revealed that in $91\%$ of cases, the LLM-generated explanations were accurate and informative, though the explanations of the remained $9\%$ were less precise than expert interpretations, including $3.5\%$ queries with \texttt{None} as output.
For the wrong cases, explanations will be corrected by experts and incorporated into the knowledge base for future retrieval, further enhancing its accuracy for subsequent queries.
To understand the impact of the number of similar vectors retrieved for augmented generation, we vary this parameter in the range from 1 to 5. Retrieving between 2 and 5 vectors showed minimal performance differences, with accuracy ranging from $89\%$ to $91\%$. However, when retrieving only 1 vector, accuracy dropped to $85\%$, and the proportion of \texttt{None} outputs increased to $8\%$. 
This result suggests that the encoding mechanism may not be perfect, but increasing the number of retrieved vectors can mitigate this limitation by leveraging the LLM's ability to adaptively contextualize information.
% Our approach using RAG also showed clear improvements over the LLM alone, with a notable increase in explanation accuracy. A detailed comparison between RAG-augmented LLM results and non-RAG results is provided in \Cref{sec:dbgpt}.
Our experiments were conducted using both Doubao and ChatGPT 4.0, and we observed minimal differences in accuracy between them. A comprehensive analysis of different language models is a future direction, along with better strategies for managing stale knowledge.

Second, we consider the end-to-end response time in our evaluation. 
This response time mainly consists of three components: the encoding overhead from the smart router, the search time within the knowledge base, and the processing and generation overhead of the LLM.
As mentioned in \Cref{sec:framework}, our smart router is lightweight and the average inference time is lower than $0.1$ ms.
Since our knowledge base is currently small (with only 20 queries), the search time per request remains under $0.1$ ms. As the this knowledge base grows, the search time will inevitably increase, but we do not expect this component to dominate, given recent advances in vector indexing\cite{malkov2018efficient}.
% This presents another promising future research direction: find a reasonable storage size that maintains high explanation accuracy while keeping retrieval times efficient.
The LLM's processing (thinking) time is generally fast ($\leq$ 2 seconds), but average generation time is around 10 seconds. This timing balance highlights that while retrieval is near-instantaneous, the generation step requires more time.
While there is room for further improvement, the end-to-end response time is acceptable because the output is meant to be consumed by users.

An additional advantage of using an LLM is its flexibility in offering a conversational interface that allows follow-up questions. For instance, in this example, a user might inquire why the predicate on the \texttt{customer} table does not benefit from the index on \texttt{c\_phone}. The LLM can provide an in-depth explanation, clarifying that many database systems cannot utilize indexes on columns when functions like \texttt{substring} are applied directly to the indexed column.

% \paragraph{Cases that LLM is worse than human}

% \paragraph{Cases that LLM is even better than human}

\subsection{Participant Study}\label{sec:study}
To evaluate users' perceptions of our explanation quality, we designed a human-subject study focused on measuring ease of understanding. To ensure a fair comparison, we divided participants equally into two groups. Both groups were given the same query, as shown in \Cref{eg:q3}, along with essential contextual information (e.g., the purpose of the survey and an overview of the hybrid engine structure).

The first group received both the AP and TP plan details (presented in JSON format for better readability) along with the LLM-generated explanation. 
We asked users to review both the plan details and the LLM explanation, record the time taken until they indicated fully understanding of the explanation. Then we ask them to submit their interpretations.
The second group initially received only the AP and TP plan details, and we similarly recorded the time they spent to fully understand the performance differences based solely on these plan details. 
Users were also asked to submit a brief description of their understanding to assess correctness. 
Then, we further provided them with the LLM-generated explanation and asked if they would like to modify or adjust their initial understanding based on this new information. Finally, we asked them to rate the difficulty of understanding both the original plan details and the LLM explanation on a scale from 0 (easiest) to 10 (hardest).
Since both groups spent time reviewing the plan details, the difference between two measured time reflects only the additional cognitive effort required to analyze and interpret the plans with or without the LLM explanation. This design ensures fairness by isolating the impact of structured output versus natural language output on user understanding.

Our results are as follows.
For participants who did not initially receive the LLM-generated explanation, 60\% correctly identified the reason for the plan performance differences between plans, with an average time of 8.2 minutes (including the time spent on reading and understanding the plan details). The remaining $40\%$ submitted incorrect reasons; however, after reviewing the LLM's generation, they were able to correct their understanding. The average difficulty rating for understanding the plan details was 8.5, while the LLM-generated explanation received an average difficulty rating of 3.
For participants who received the LLM-generated results from the start, the average time taken to understand the reason was 3.5 minutes, and all users in this group were able to summarize the correct reason. These results indicate that the LLM-generated explanation reduces both the time required for understanding and the perceived difficulty, enhancing user comprehension of query performance differences.

\normalsize
\subsection{Comparison with Other Methods}\label{sec:dbgpt}
We additionally compare our approach with DBG-PT\cite{DBG-PT}, which aims to suggesting hints for debugging regressions in query execution time. DBG-PT leverages LLMs to identify and reason about structural differences between query plans from the same engine.
The generated results for the query in \Cref{eg:q3} are present in \Cref{tab:explain}. DBG-PT performs well in analyzing structured plans by comparing their differences in detail. For the input of DBG-PT, we only provide the TP and AP plan details without any historical query or expert explanation. For a fair comparison, we adjusted the prompts in our method by removing RAG-related context but retained the same plan details and any additional user prompts provided to the LLM. 
The results indicate that while DBG-PT can effectively interpret plan details, it often contains errors in its explanations. After testing additional queries from our dataset and reviewing the explanations, we identified the following key limitations:
\begin{itemize}
    \item \textit{Fundamental errors:} It may misinterpret index usage. For instance, when a query includes a predicate like \texttt{substring(c\_phone, 1, 2) in (...)}, no index is used; however, it still assumes AP is faster due to perceived index benefits.

    \item \textit{Overemphasis on minor factors:} DBG-PT often overemphasizes column-oriented storage as the key reason for AP's speed, while underemphasizing critical factors such as TP's lack of indexes or inefficient join methods.

    \item \textit{Ignoring limitations:} Despite instructions to avoid comparing costs between AP and TP, DBG-PT still seems to rely on cost differences sometimes, which is problematic because these costs are calculated differently and do not correlate well with real execution latencies.

    \item \textit{Lack of context for relative values:} DBG-PT struggles to assess the significance of certain values without experience. For example, it cannot determine whether the size of an \texttt{OFFSET} or \texttt{LIMIT} is large enough to impact plan efficiency without historical execution data.
\end{itemize}
% \normalsize
% \textcolor{red}{
% \paragraph{ when LLM is bad?} (e.g. long string) and the observation is much different with previous knowledge. }

\section{Conclusion and future work}\label{sec:future}
In conclusion, our study demonstrates the effectiveness of a RAG-augmented LLM framework in providing user-friendly, accurate explanations for query performance in HTAP systems. By leveraging expert knowledge through a knowledge base containing past explanations, our approach enhances the accuracy and relevance of LLM-generated explanations, allowing users to better understand complex execution plans without needing specialized expertise. This framework effectively balances efficiency and accuracy by utilizing pre-trained public models with targeted context augmentation, reducing reliance on costly expert intervention.

Several areas remain open for future work, including developing strategies for maintaining the knowledge base (including selecting representative queries and expiring stale queries), establishing benchmarks to measure accuracy improvements, and finally working towards general explanations for the question ``Why does my query run so slowly?''.
The promising results from our work demonstrate the potential of RAG-augmented LLM in enabling more reliable, scalable, and intelligent solutions for automated database performance analysis.

\bibliographystyle{plain}
\bibliography{sample}

\begin{thebibliography}{10}

\bibitem{achiam2023gpt}
Josh Achiam, Steven Adler, Sandhini Agarwal, Lama Ahmad, Ilge Akkaya, Florencia~Leoni Aleman, Diogo Almeida, Janko Altenschmidt, Sam Altman, Shyamal Anadkat, et~al.
\newblock Gpt-4 technical report.
\newblock {\em arXiv preprint arXiv:2303.08774}, 2023.

\bibitem{bai2022constitutional}
Yuntao Bai, Saurav Kadavath, Sandipan Kundu, Amanda Askell, Jackson Kernion, Andy Jones, Anna Chen, Anna Goldie, Azalia Mirhoseini, Cameron McKinnon, et~al.
\newblock Constitutional ai: Harmlessness from ai feedback.
\newblock {\em arXiv preprint arXiv:2212.08073}, 2022.

\bibitem{Doubao}
ByteDance.
\newblock Doubao large language model, https://www.doubao.com/, 2024.

\bibitem{chen2022bytehtap}
Jianjun Chen, Yonghua Ding, Ye~Liu, Fangshi Li, Li~Zhang, Mingyi Zhang, Kui Wei, Lixun Cao, Dan Zou, Yang Liu, et~al.
\newblock Bytehtap: bytedance's htap system with high data freshness and strong data consistency.
\newblock {\em Proceedings of the VLDB Endowment}, 15(12), 2022.

\bibitem{chen2024bytehtap}
Jianjun Chen, Li~Zhang, et~al.
\newblock Vedb-htap: a highly integrated, efficient and adaptive htap system.
\newblock {\em \url{https://github.com/Hap-Hugh/Query-performance-explanation/blob/main/vedb-htap.pdf} \\ This paper was under submission. We will correct the citation when it becomes public.}

\bibitem{rag2023}
Yunfan Gao, Yun Xiong, Xinyu Gao, Kangxiang Jia, Jinliu Pan, Yuxi Bi, Yi~Dai, Jiawei Sun, Meng Wang, and Haofen Wang.
\newblock Retrieval-augmented generation for large language models: A survey.
\newblock {\em arXiv preprint arXiv:2312.10997}, 2023.

\bibitem{DBG-PT}
Victor Giannakouris and Immanuel Trummer.
\newblock Dbg-pt: A large language model assisted query performance regression debugger.
\newblock {\em Proceedings of the VLDB Endowment}, 17(12), 2024.

\bibitem{rag2020}
Patrick Lewis, Ethan Perez, Aleksandra Piktus, Fabio Petroni, Vladimir Karpukhin, Naman Goyal, Heinrich K{\"u}ttler, Mike Lewis, Wen-tau Yih, Tim Rockt{\"a}schel, et~al.
\newblock Retrieval-augmented generation for knowledge-intensive nlp tasks.
\newblock {\em Advances in Neural Information Processing Systems}, 33:9459--9474, 2020.

\bibitem{lewis2020retrieval}
Patrick Lewis, Ethan Perez, Aleksandra Piktus, Fabio Petroni, Vladimir Karpukhin, Naman Goyal, Heinrich K{\"u}ttler, Mike Lewis, Wen-tau Yih, Tim Rockt{\"a}schel, et~al.
\newblock Retrieval-augmented generation for knowledge-intensive nlp tasks.
\newblock {\em Advances in Neural Information Processing Systems}, 33:9459--9474, 2020.

\bibitem{malkov2018efficient}
Yu~A Malkov and Dmitry~A Yashunin.
\newblock Efficient and robust approximate nearest neighbor search using hierarchical navigable small world graphs.
\newblock {\em IEEE transactions on pattern analysis and machine intelligence}, 42(4):824--836, 2018.

\bibitem{marcus2021bao}
Ryan Marcus, Parimarjan Negi, Hongzi Mao, Nesime Tatbul, Mohammad Alizadeh, and Tim Kraska.
\newblock Bao: Making learned query optimization practical.
\newblock In {\em Proceedings of the 2021 International Conference on Management of Data}, pages 1275--1288, 2021.

\bibitem{singh2024panda}
Vikramank Singh, Kapil~Eknath Vaidya, Vinayshekhar~Bannihatti Kumar, Sopan Khosla, Murali Narayanaswamy, Rashmi Gangadharaiah, and Tim Kraska.
\newblock Panda: Performance debugging for databases using llm agents.
\newblock 2024.

\bibitem{touvron2023llama}
Hugo Touvron, Louis Martin, Kevin Stone, Peter Albert, Amjad Almahairi, Yasmine Babaei, Nikolay Bashlykov, Soumya Batra, Prajjwal Bhargava, Shruti Bhosale, et~al.
\newblock Llama 2: Open foundation and fine-tuned chat models.
\newblock {\em arXiv preprint arXiv:2307.09288}, 2023.

\bibitem{xu2023coool}
Xianghong Xu, Zhibing Zhao, Tieying Zhang, Rong Kang, Luming Sun, and Jianjun Chen.
\newblock Coool: A learning-to-rank approach for sql hint recommendations.
\newblock {\em 5th International Workshop on Applied AI for Database Systems and Applications}, 2023.

\bibitem{zhou2023d}
Xuanhe Zhou, Guoliang Li, Zhaoyan Sun, Zhiyuan Liu, Weize Chen, Jianming Wu, Jiesi Liu, Ruohang Feng, and Guoyang Zeng.
\newblock D-bot: Database diagnosis system using large language models.
\newblock {\em arXiv preprint arXiv:2312.01454}, 2023.

\bibitem{zhou2024llm}
Xuanhe Zhou, Xinyang Zhao, and Guoliang Li.
\newblock Llmfor data management.
\newblock {\em Proceedings of the VLDB Endowment}, 17(12), 2024.

\bibitem{zhu2023lero}
Rong Zhu, Wei Chen, Bolin Ding, Xingguang Chen, Andreas Pfadler, Ziniu Wu, and Jingren Zhou.
\newblock Lero: A learning-to-rank query optimizer.
\newblock {\em Proceedings of the VLDB Endowment}, 16(6):1466--1479, 2023.

\end{thebibliography}
\end{document}